# Design and Implementation of an Improved Carry Increment Adder

Aribam Balarampyari Devi[1], Manoj Kumar[2] and Romesh Laishram[3]

[1] M. Tech VLSI & Embedded System, National Institute of Technology, Manipur, India.
`balaramaribam@gmail.com`
[2]ECE Department, National Institute of Technology, Manipur, India.
`manojara400@gmail.com`
[3]ECE Department, Manipur Institute of Technology, Manipur, India.
`romeshlaishram@gmail.com`

## ABSTRACT

*A complex digital circuit comprises of adder as a basic unit. The performance of the circuit depends on the design of this basic adder unit. The speed of operation of a circuit is one of the important performance criteria of many digital circuits which ultimately depends on the delay of the basic adder unit. Many research works have been devoted in improving the delay of the adder circuit. In this paper we have proposed an improved carry increment adder (CIA) that improves the delay performance of the circuit. The improvement is achieved by incorporating carry look adder (CLA) in the design of CIA contrary to the previous design of CIA that employs ripple carry adder (RCA). A simulation study is carried out for comparative analysis. The coding is done in Verilog hardware description language (HDL) and the simulation is carried out in Xilinx ISE 13.1 environment.*

## 1. INTRODUCTION

The demand of high-performance VLSI systems are increasingly rapidly  for used in small and portable devices, standard wireless and mobile devices, and rapidly advancing biomedical instruments [1], [2]. The performance of a VLSI systems are basically decided by the speed of operation, low power consumption and less design area. In the recent years due to the fast growing technologies in mobile communication and computation, the demand for building low-power VLSI systems has also increases. The technology of storage batteries does not advance at the same rate as the microelectronics technology. Therefore the power available for mobile systems is a very limited. So system designers have to comply with more constraints such as high speed, high throughput, small silicon area, and at the same time, low-power consumption [1].A complex digital system consists of many operational blocks and arithmetic logic unit (ALU) is one of the important building block of such system. The main component of an ALU involves several binary adders. Eventually an efficient adder design essentially improves the performance of a complex digital system. The design of area and power-efficient high-speed data  path logic systems is one of the most substantial areas of research in VLSI system design. In binary adders, the speed of operation is limited by the time taken in propagating the carry through the adder. The sum for each bit position in a basic adder is obtained sequentially only after the previous bit position has been calculated and a carry propagated into the next position. The development of adder has seen tremendous growth since the design of basic half adder and full adder. Many works have been dedicated to improve the ripple carry adder in terms of delay ,area, power consumption etc. Since adders are used as modules in many complex digital circuit, improving the performance of the digital adder would extensively speed up the execution of binary operations inside such complex circuit that comprised of many adder blocks. In cell-based VLSI design techniques the parameters may be well be characterized in terms of circuit area and speed as well as suitability for logic optimization and synthesis[3]. Over the last decade many different adder architectures for improving the binary addition have been design and proposed. In this paper our main focus is to improve the delay performance of the binary adder circuit thereby increasing the speed of operation.

The rest of the paper is organized as follows. Section 2 provides a brief review of the related work done earlier on the design binary adders. In Section 3, the design of carry increment and modified or improved carry increment adder is presented. In Section 4, analysis on the simulation results is done and the conclusion is drawn in Section 5.

## 2. RELATED WORKS

Some of the binary adders are summarized in this section. The most basic simple adder is the Ripple Carry Adder (RCA)[3][4] but it is the slowest with $O(n)$ area and $O(n)$ delay, where n is the operand size in bits. Carry Look-Ahead (CLA)[5][6] have $O(n \cdot \log(n))$ area and $O(\log(n))$ delay, but typically suffer from irregular layout. An area efficient CLA is proposed by the authors in [7]. On the other hand, Carry Skip Adder(CSA)[8][9],carry increment adder (CIA) [6],[11]and carry select adder (CSLA)[10], [12]-[16] provides a good compromise in terms of area and delay, along with a simple and regular layout. Carry save adder have O(n) area and $O(\log n)$ delay. CLA adders can be realized in two gate levels provided there is no limit on fan in/out. CSLA reduces the computation time by pre-computing the sum for all possible carry bit values (i.e. '0' and '1'). After the carry becomes available the correct sum is selected using multiplexer. A power efficient adder scheme is proposed in [17]. It is shown by the authors in [6],[18],[19] that among the parallel adders carry increment adder has the best delay performance which is one of the most important parameter in the high speed devices. This result prompted us to study the details of CIA and an a modified architecture of CIA is proposed in this paper that further improves the delay of the circuit. The architecture incorporates carry look ahead adder. The proposed design is a new concept and to the best of our knowledge it has not been proposed earlier by any researchers. The details of the design are explained in the next subsequent section.

## 3. CARRY INCREMENT ADDER (CIA)

In this section the design and working of CIA is presented. First the conventional design of CIA which is based on RCA is elaborated. We named this design of CIA as CIA_RCA. Secondly the modified CIA is presented and we named it as CIA_CLA.

### 3.1. CIA_RCA

The standard Carry Increment Adder (CIA) consists of RCA's and incremental circuitry [18]. The incremental circuit is designed using HA's in ripple carry chain with a sequential order. The addition operation is done by separating the total number of bits in to group of 4bits and addition operation is performed by several 4-bit RCA's. Instead of computing two partial sums for each group and selecting the correct one, only one partial sum is calculated and incremented if necessary, according to the input carry. Thus the second adder and the multiplexers in the carry-select scheme can be replaced by a much smaller incremental circuit and the modified architecture is the Carry Increment Adder (CIA)[19]. For example, an 8-bit CIA comprises of two 4-bit RCA. The first block of RCA adds first 4-bits to produce 4-bit partial sum and a carry output. Thus, first 4-bit of sum of CIA is directly obtained from first block of RCA. And the carry output of first RCA block is given as input to the $c_{in}$ of incremental circuit. Incremental circuit consists of Half Adders (HA). Hence, the partial sum obtained from the second RCA block is given to incremental circuit. The block diagram representation of an 8-bit CIA_RCA is as shown in Figure 1.

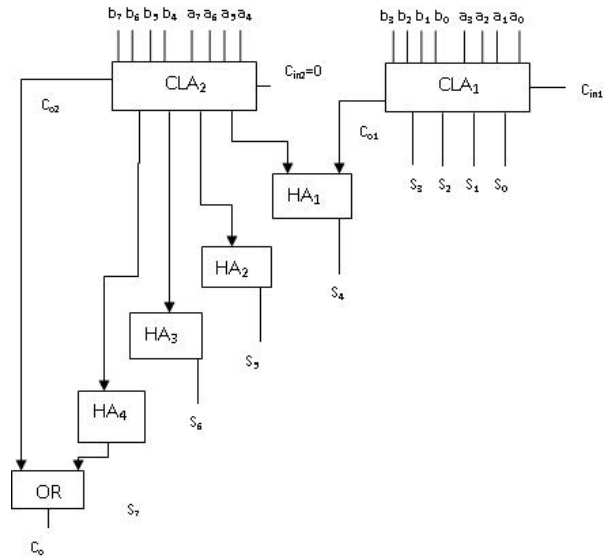

Figure 1. Block diagram of CIA_RCA

## 3.2. CIA_CLA

In this subsection we present the modified carry increment adder i.e. CIA_CLA. We know that RCA is the basic binary adder circuit and is quite popular because of its simple design. However it suffers from the worst propagation delay affecting the overall performance of the system. It is proved that CLA performs better than RCA in terms of delay at the expense of increased design complexity. We have modified CIA_RCA by replacing the RCA with CLA block. It is quite obvious because of the property of CLA, the overall delay performance will be improved. As similar to CIA_RCA incremental circuit can be designed using HA's in ripple carry chain with a sequential order. The block diagram representation of CIA_CLA is as shown in Figure 2.

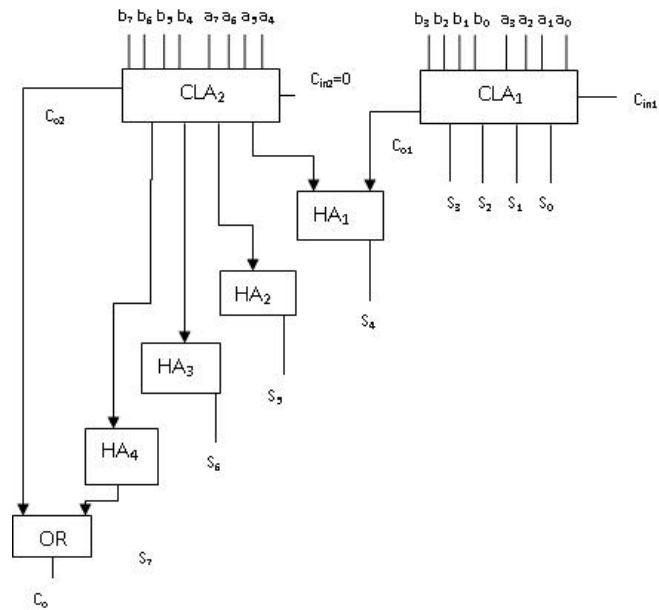

Figure 2. Block diagram of CIA_CLA

## 4. SIMULATION RESULTS

The simulation study of the proposed CIA_CLA is presented in this section. The RTL code is written using Verilog HDL and the simulation and design synthesis is carried under Xilinx ISE 13.1 environment. The behavioural simulation result for verifying the design is done under behavioural simulation by writing a test bench. The RTL schematic and technology view are also generated using the synthesis tool. We have performed the experiment for addition of 8 bit binary numbers. The CIA_RCA is built using two modules of 4-bit RCA and the incremental circuitry by the half adders. Whereas CIA_CLA is built using two modules 4-bit CLA. The delay of 4-bit RCA is about 12.008 ns but the delay of 4-bit CLA is about 10.006 ns. This indicates that CLA is faster than RCA. Therefore it is quite obvious that CIA_RCA will be better than CIA_RCA in terms of delay performance. The behavioural simulation result, RTL schematic, and technology view schematic of CIA_RCA and CIA_CLA are shown in Figure 3 to Figure 8 respectively.

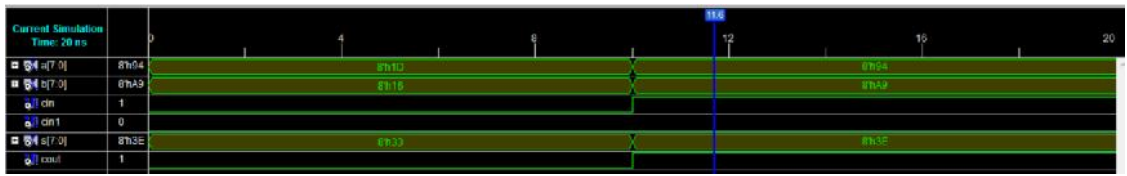

Figure 3. Simulation result of two 8-bit addition using CIA_RCA

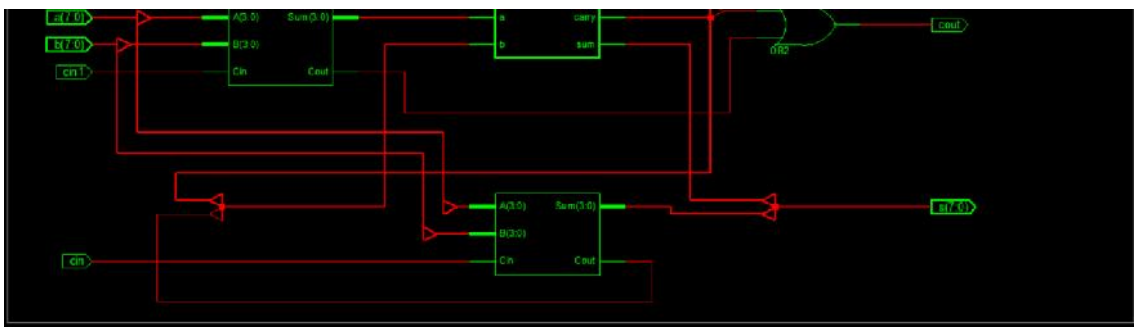

Figure 4. RTL schematic view of CIA_RCA

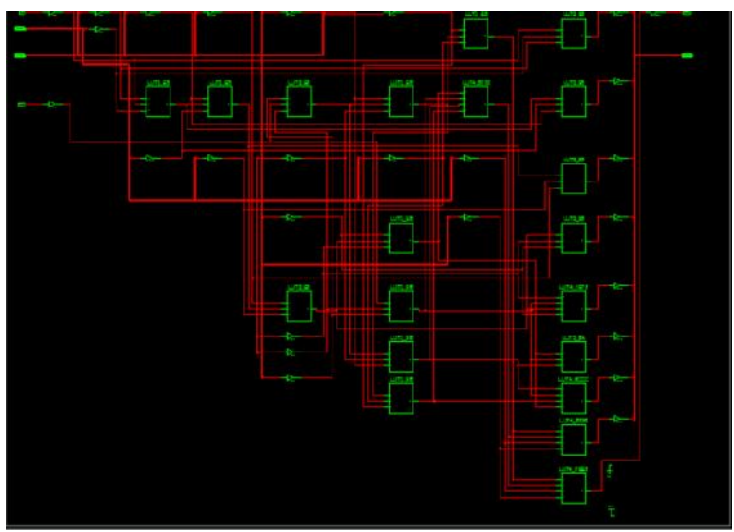

Figure 5. Technology view of CIA_RCA

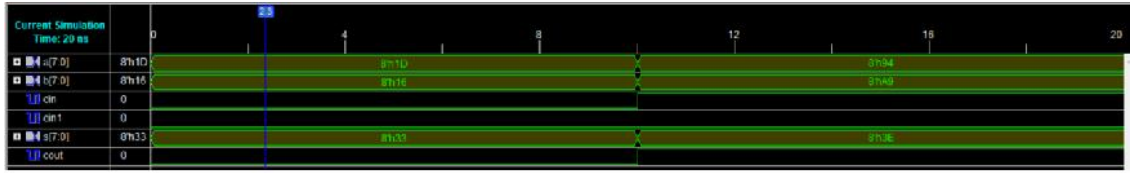

Figure 6. Simulation result of two 8-bit addition using CIA_CLA

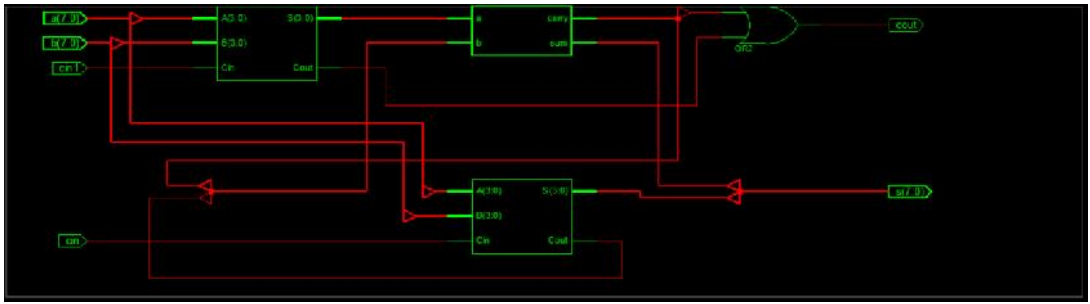

Figure 7. RTL schematic view of CIA_CLA

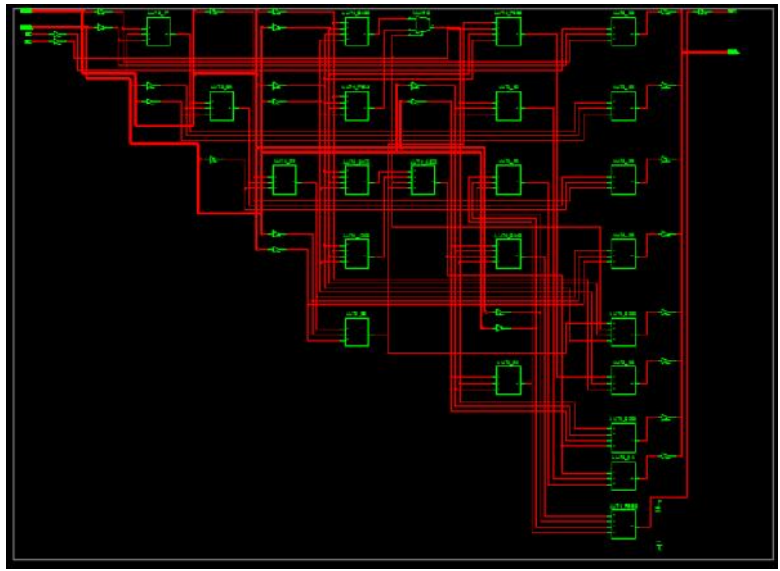

Figure 8. Technology view of CIA_CLA

The comparison of CIA_RCA and CIA_CLA in terms delay, area and power consumption are given in Table 1. The performance parameters like delay and area are obtained from the synthesis report and the power dissipation or consumption is calculated using x-power analyzer which is available with the Xilinx ISE design suite. It can be observed form Table 1 that the proposed design has better delay performance which is the desired goal of this research without compromising the power dissipation.

Table 1. Comparison of 8 bit CIA_RCA and CIA_CLA.

| Design | Area(LUT's) | Area(Slices) | Delay(ns) | Power consumption (mW) |
|---|---|---|---|---|
| CIA_RCA | 20 | 13 | 14.59 | 41 |
| CIA_CLA | 19 | 12 | 13.54 | 41 |

## 5. CONCLUSIONS

A modified carry increment adder is proposed in this paper using the faster carry look ahead modules instead of the much slower ripple carry adder. By replacing the 4-bit RCA with a 4-bit CLA the delay performance of the circuit is improved in the design without affecting the poer dissipation of the circuit. The design is tested and verified by Verilog HDL coding and simulation is carried out in Xilinx ISE 13.1 environment. In this paper we have carried out the design only for two 8-bit binary addition but the design may be extended to higher order adder circuit. It may pointed out that although CLA has better delay performance than RCA the circuit complexity of CLA increases as the number of bit increases. The choice of a particular adder depends on the intended application. Future work may be dedicated to study the complexity of CIA_CLA when the number of bit is increased.

**AUTHORS**

**Aribam Balarampyari Devi** is presently doing M.Tech. in VLSI & Embedded systems in NIT, Manipur. Her research area includes digital sytems, Circuit design, VLSI design.

**Manoj Kumar** is presently working as Assistant professor in the department of Electronics & Communication Engineering, National Institute of Technology, Manipur-India. His research area includes digital electronics, VLSI design.

**Romesh Laishram** is presently working as Assistant professor in the department of Electronics & Communication Engineering, Manipur Institute of Technology, Imphal-India. He has published more than 20 research papers in refereed journals and conferences. He has also publish one book on "WiMAX: Call Admission Control". He is also a member of the editorial board in International Journal of Recent Advances in Engineering & Technology(IJRAET). His research area includes signal processing, image processing, Communication systems, VLSI design.